\documentclass{emulateapj}
\usepackage{xspace}
\usepackage{amsmath}
\usepackage{hyperref}
\usepackage{framed} 
\usepackage{txfonts}
\usepackage{tikz}
\newcommand*\circled[1]{\tikz[baseline=(char.base)]{\node[shape=circle,draw,inner sep=1pt] (char) {#1};}}

\def\dim#1{\mbox{\,#1}}
\def\hide#1{}

\begin{document}

\title{On the Proper Use of the Reduced Speed of Light Approximation}

\author{Nickolay Y.\ Gnedin\altaffilmark{1,2,3}}
\altaffiltext{1}{Particle Astrophysics Center, Fermi National Accelerator Laboratory, Batavia, IL 60510, USA; gnedin@fnal.gov}
\altaffiltext{2}{Kavli Institute for Cosmological Physics, The University of Chicago, Chicago, IL 60637 USA;}
\altaffiltext{3}{Department of Astronomy \& Astrophysics, The
  University of Chicago, Chicago, IL 60637 USA} 

\begin{abstract}
I show that the Reduced Speed of Light (RSL) approximation, when used properly (i.e.\ as originally designed - \emph{only for the local sources but not for the cosmic background}), remains a highly accurate numerical method for modeling cosmic reionization. Simulated ionization and star formation histories from the ``Cosmic Reionization On Computers'' (CROC) project are insensitive to the adopted value of the reduced speed of light for as long as that value does not fall below about 10\% of the true speed of light. A recent claim of the failure of the RSL approximation in the Illustris reionization model appears to be due to the effective speed of light being reduced in the equation for the cosmic background too, and, hence, illustrates the importance of maintaining the correct speed of light in modeling the cosmic background.
\end{abstract}


\maketitle

The primary challenge of simulating radiative transfer in astrophysics is in the high value of the speed of light - the high dimensionality of the problem is a less severe technical challenge, since dynamics of dark matter (``N-body'') is also six-dimensional, but from a technical point of view it is a solved problem.

One of way to cope algorithmically with the extremely high value for the speed of light is a Reduced Speed of Light (RSL) approximation \citep{ng:ga01}. The idea behind the RSL approximation is simple - since most of astrophysical dynamics is modeled in the Newtonian limit (i.e.\ the leading term in the Taylor series expansion over powers of $1/c$) anyway, it is only important that the higher order terms be small compared to the leading one. For a system with the characteristic velocity $v$ the subsequent terms are of the order of $v/c$, and as long as $v/c$ is much less than unity, the Newtonian limit is valid. Hence, it does not matter which value for $c$ to take as long as $v$ remains much less than the modified, ``reduced'' value, which I will label $\hat{c}$ hereafter.

One, of course, has to be careful, because the idea presented in the previous paragraph only applies to dynamics of nonrelativistic matter, and there are many other processes in physics, including the dynamics of photons themselves, where the specific value of $c$  actually matters. Unfortunately, occasionally this concept is being confused and the RSL approximation is used incorrectly. Hence, the purpose of this short paper is to clarify when one can and when one cannot use the RSL approximation in cosmological simulations.

It is instructive to start with the cosmological radiative transfer equation for the monochromatic radiation energy density $I_\nu(t,\vec{x})$ (measured in ergs per cm$^3$ per hertz) as a function of cosmic time $t$ and comoving position $\vec{x}$,
\begin{equation}
  \frac{\partial I_\nu}{\partial t} + H\left(\nu \frac{\partial I_\nu}{\partial \nu}-3I_\nu\right) + \vec{n}\frac{c}{a}\frac{\partial I_\nu}{\partial\vec{x}} = -\kappa_\nu I_\nu + S_\nu,
  \label{eq:rt}
\end{equation}
where $a$ is the cosmological scale factor, $H$ is the Hubble parameter, $\kappa_\nu$ is the absorption coefficient (per unit time), and $S_\nu$ is the source function. The absorption coefficient is usually a sum over various absorption processes,
\[
  \kappa_\nu = c \sum_j \sigma_j n_j,
\]
where $n_j$ is the number density of some absorbing species $j$ and $\sigma_j$ is the cross-section of the appropriate atomic process; both of them are not affected by the RSL approximation in any way. 

In order to introduce the RSL approximation as it was originally designed, it is instructive to split the full radiation energy density into two components: the mean cosmic background $\bar{I}_\nu(t) \equiv \langle I_\nu\rangle_V$, and the fluctuation around the mean $\delta I_\nu(t,\vec{x}) \equiv I_\nu - \bar{I}_\nu$. Equation for the cosmic background is easily derivable by spatially averaging equation (\ref{eq:rt}),
\begin{equation}
  \frac{\partial \bar{I}_\nu}{\partial t} + H\left(\nu \frac{\partial \bar{I}_\nu}{\partial \nu}-3I_\nu\right) = -\bar{\kappa}_\nu \bar{I}_\nu + \bar{S}_\nu,
  \label{eq:bkgr}
\end{equation}
where the mean absorption coefficient $\bar{\kappa}_\nu \equiv \langle \kappa_\nu I_\nu\rangle/\bar{I}_\nu$ is radiation energy density weighted.

\begin{figure*}[th]
\includegraphics[width=0.5\hsize]{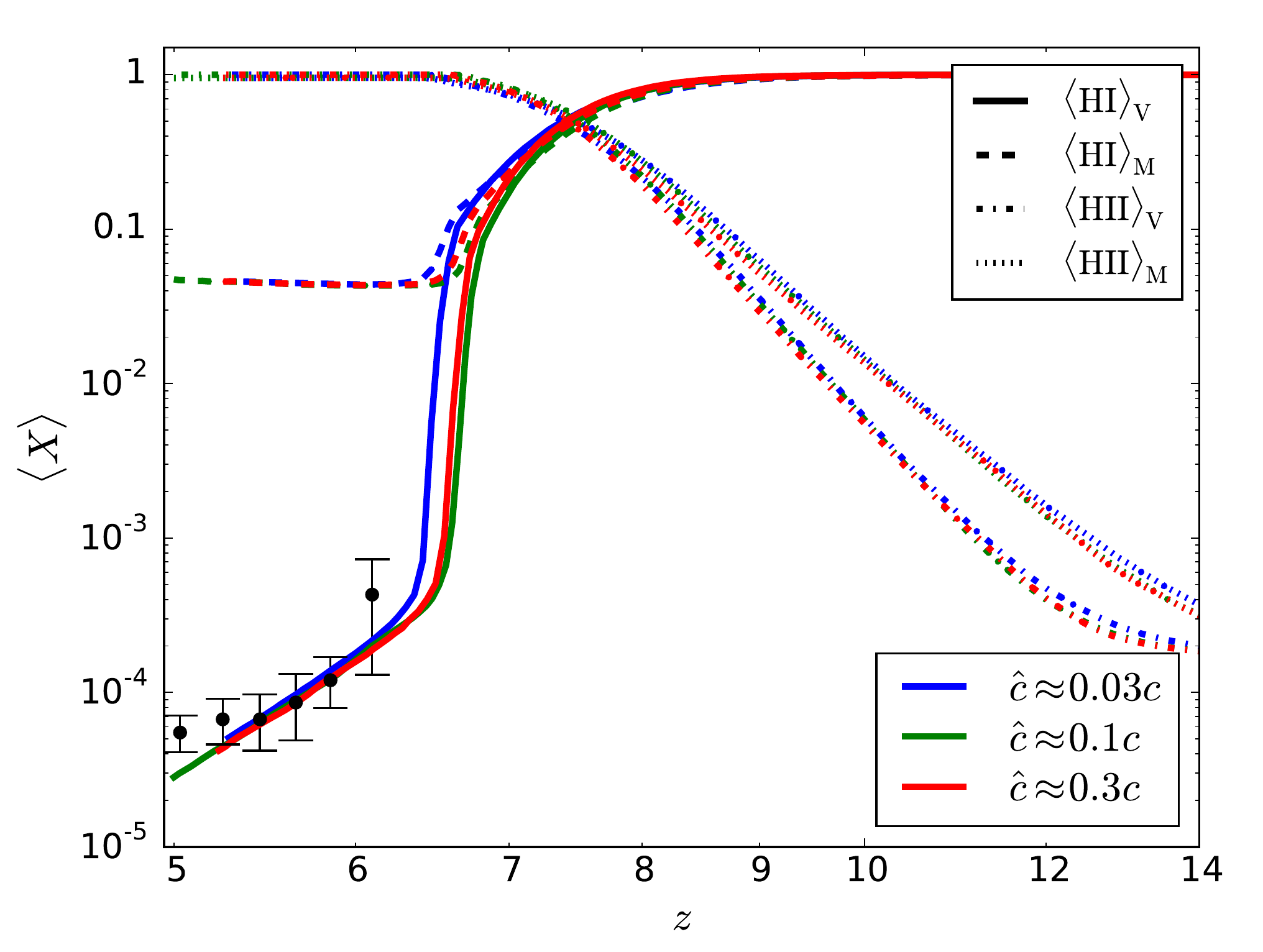}%
\includegraphics[width=0.5\hsize]{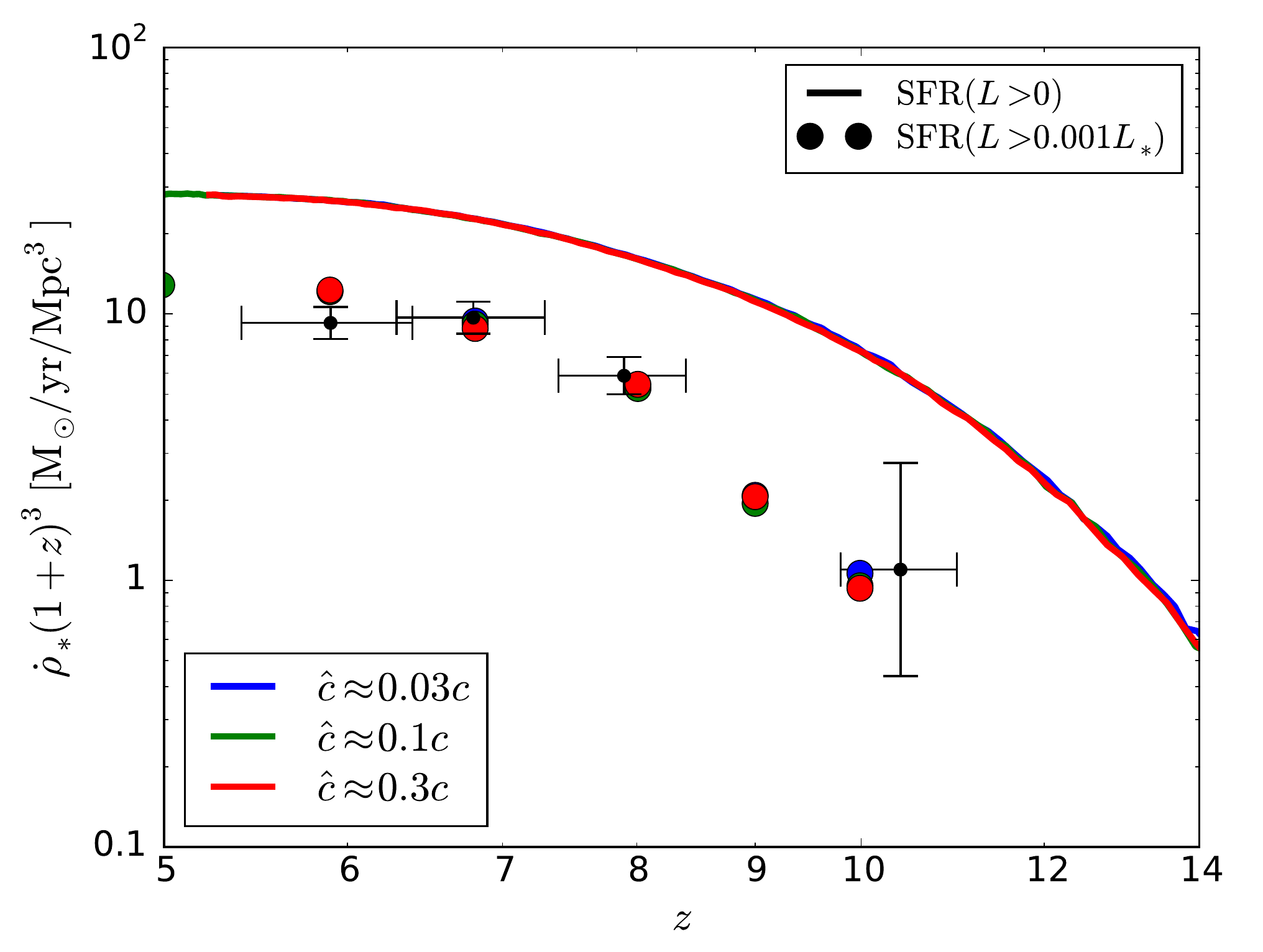}%
\caption{Ionization (left) and star formation (right) histories for 3 CROC simulations with varied effective speed of light $\hat{c}$. In the right panel both the true global star formation rate and star formation rate in galaxies above $0.001 L_*$ are shown; data points are from \citet{rei:biot15}. For $\hat{c}\la 0.1c$ the effect of the reduced speed of light is barely noticeable.\label{fig:rsl}}
\end{figure*}

In some circumstances the equation for the fluctuations in the radiation energy density can be simplified. For example, when modeling cosmic reionization (the actual specific application considered in this paper), and while restricting radiation under consideration to ionizing only (mean free path much shorter than the cosmic horizon), cosmological expansion and redshift can be neglected over the time a photon crosses the mean free path distance. Hence, terms proportional to the Hubble parameter can be omitted. Equation for the fluctuation then becomes (after multiplying by $a/c$),
\begin{equation}
  \frac{a}{c}\frac{\partial }{\partial t}\delta I_\nu + \vec{n}\frac{\partial }{\partial\vec{x}}\delta I_\nu = - \frac{1}{\lambda_\nu} \delta I_\nu + \psi_\nu,
  \label{eq:drt}
\end{equation}
where ${\lambda_\nu}(t,\vec{x})$ is the local \emph{comoving} photon mean free path,
\[
  \frac{1}{\lambda_\nu} \equiv a \sum_j \sigma_j n_j,
\]
which is independent of the speed of light, and a new source function $\psi_\nu$ is defined as
\begin{equation}
  \psi_\nu = \frac{a}{c}\left[ \sum_k L_k (n_k-\bar{n}_k) + (\bar{\kappa}_\nu-\kappa_\nu)\bar{I}_\nu\right],
  \label{eq:psi}
\end{equation}
where the sum is over all sources with luminosities $L_k$ and number densities $n_k$ (galaxies of different masses, quasars of different luminosities, etc).

It is only equation (\ref{eq:drt}) that can be solved in a Newtonian limit in some circumstances; namely, if all the \emph{sources evolve on timescales longer than the light crossing time of the photon mean free path} and all cosmic \emph{ionization fronts move much slower than the speed of light}. 

In this case one can introduce the Reduced Speed of Light approximation by replacing $c$ in the first term in equation (\ref{eq:drt}) with the effective speed of light $\hat{c}$ (circled),
\begin{equation}
  \frac{a}{\circled{$\hat{c}$}}\frac{\partial }{\partial t}\delta I_\nu + \vec{n}\frac{\partial }{\partial\vec{x}}\delta I_\nu = - \frac{1}{\lambda_\nu} \delta I_\nu + \psi_\nu.
  \label{eq:rsl}
\end{equation}
It is important to emphasize that this is the \emph{only} place where the speed of light should be reduced. In particular, the speed of light should \emph{not} be reduced in the equation for the cosmic background, as that would make background evolution incorrect, and is also not needed, since it is trivial to solve equation(\ref{eq:bkgr}). Nor should $c$ that enters the definition for $\psi_\nu$ in equation (\ref{eq:psi}) be reduced, as it would result in the incorrect photon production rate.


In order to illustrate how the RSL approximation perform when used properly, I use Cosmic Reionization On Computers (CROC) simulations of cosmic reionization \citep{ng:g14}. CROC simulations are performed with the Adaptive Refinement Tree (ART) code \citep{misc:k99,misc:kkh02,sims:rzk08}. In the ART code equation (\ref{eq:rsl}) is implemented in a further transformed form and is solved with the Optically Thin Variable Eddington Tensor (OTVET) method of \citet{ng:ga01}. The complete description of the CROC radiative transfer solver, down to finite difference operator and accuracy tests, is presented in the Appendix C of \citet{ng:g14}.

In particular, the ART implementation of the RSL approximation does not adopt any specific value for $\hat{c}$, but instead imposes a fixed ratio of the hydrodynamic timestep $\Delta t_{\rm H}$ and the radiative transfer timestep $\Delta t_{\rm RT}$,
\[
  \Delta t_{\rm RT} = \frac{\Delta t_{\rm H}}{N_{\rm RT}},
\]
where $N_{\rm RT}$ is the number of time the radiative transfer solver is ``subcycled'' (i.e.\ makes a timestep) for one hydrodynamic timestep. The hydrodynamic timestep is set by the hydrodynamic Courant-Friedrichs-Lewy condition,
\[
   \Delta t_{\rm H} = C_{\rm CFL} \frac{\Delta r}{v_{\rm MAX}},
\]
where $C_{\rm CFL}=0.5$ is the Courant-Friedrichs-Lewy number (a property of the hydrodynamic solver), $\Delta r$ is the spatial resolution, and $v_{\rm MAX}$ is the maximum total (i.e.\ bulk plus sound) velocity on the grid (for an AMR code this condition is more complicated, accounting for different cells sizes at different refinement levels, but conceptually it is equivalent to a simple uniform grid). The radiative transfer solver sets its timestep as
\[
   \Delta t_{\rm RT} = \frac{\Delta r}{\hat{c}},
\]
hence in CROC simulations there is a relationship between $\hat{c}$ and the true speed of light,
\[
  \hat{c} = \left(\frac{N_{\rm RT}}{C_{\rm CFL}}\right) \left(\frac{v_{\rm MAX}}{c}\right) c.
\]
Typically in the simulations with the box size of $20h^{-1}\dim{Mpc}$ (used in this paper for testing) during the peak of reionization $v_{\rm MAX}\approx 500\dim{km/s}$, and I used $N_{\rm RT}=30$ as the fiducial number \citep[based on tests presented in][]{ng:g14}, so in the CROC production runs $\hat{c} \approx 0.1c$. The value of $\hat{c}$ gradually increases as the simulation proceeds, since gravitational clustering and stellar feedback drive gas to progressively higher velocities. It is also higher in larger box simulations, which includes more massive galaxies with higher escape velocities. By varying $N_{\rm RT}$ a different ratio of $\hat{c}/c$ can be implemented in the simulations.

Figure \ref{fig:rsl} demonstrate the accuracy of the RSL approximation, as implemented in the CROC simulations. Two panels show ionization histories and star formation histories for three runs, all in $20h^{-1}\dim{Mpc}$ boxes (larger box sizes would be too expensive for a such purely technical test), with the effective speed of light $\hat{c}$ set to approximately 3, 10, and 30\% of $c$ ($N_{\rm RT}=10$, 30, and 100 respectively). Clearly, production CROC simulations (with $\hat{c}\ga 0.1c$) are not compromised by the use of the RSL approximation.

This conclusion stands in conflict with the recent claim by \citet{newrei:bsv15}, who found a large difference in the ionization history of the Illustris simulation when changing the effective speed of light from $c$ to $0.1c$. At face value, that result is surprising - such a change would imply that most of ionization fronts in the Illustris simulation propagate much faster than $0.1c$ (otherwise, there would not be any difference). This is in conflict with most of studies of reionization, which find that reionization proceeds over a range of redshifts, with duration comparable to the Hubble time $t_H$. In the latter case the typical speed of ionization fronts would be $R_B/t_H = (R_B/R_H) c \ll c$, where $R_B$ is the typical size of an ionized bubble and $R_H$ is the Hubble radius. At $z=6$ the Hubble radius $R_H\sim 3{,}000\dim{Mpc}$ in comoving units, so for $R_B\sim30\dim{Mpc}$ (around the largest comoving size fitting into the Illustris simulation volume) the ionization front would move with about $3{,}000\dim{km/s}$, a speed typically found in many reionization simulations.

\begin{figure}[t]
\includegraphics[width=\hsize]{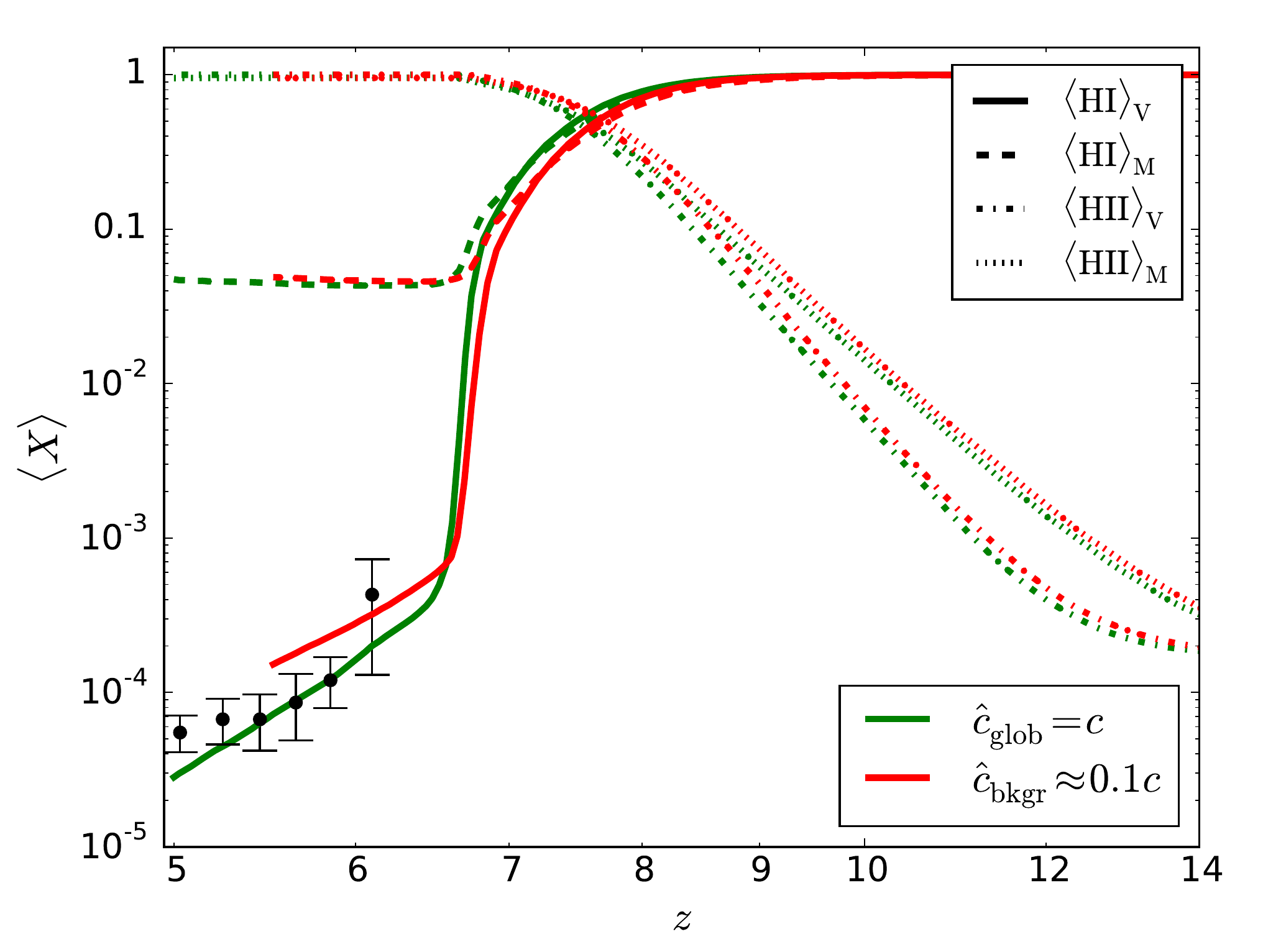}%
\caption{Ionization histories for two test simulations with various implementations of the RSL approximation: the correct one (green - the same line as in figure \ref{fig:rsl})) and another implementation with the speed of light also reduced (incorrectly) in the background equation (red).\label{fig:err}}
\end{figure}

In order to explore the potential reasons for that discrepancy, I have also implemented a version of the RSL approximation, which also modifies the equation for the cosmic background. Specifically, rewriting equation (\ref{eq:bkgr}) with the RHS in a form similar to equation (\ref{eq:rsl}),
\begin{equation}
\frac{a}{\circled{$c$}} \frac{\partial \bar{I}_\nu}{\partial t} + \frac{a}{\circled{$c$}}H\left(\nu \frac{\partial \bar{I}_\nu}{\partial \nu}-3I_\nu\right) = -\frac{1}{\bar{\lambda}_\nu} \bar{I}_\nu + \bar{\psi}_\nu,
  \label{eq:bmod}
\end{equation}
I replace both circled $c$ with $0.1c$ similarly to how the RSL approximation is used in equation (\ref{eq:rsl} (labeled $\hat{c}_{\rm bkgr}=0.1c$). The ionization history for the so modified background equation is shown in figure \ref{fig:err} (red line) together with the simulation with the correctly implemented RSL approximation (green line, the same as in figure \ref{fig:rsl}). Such a (improperly) modified RSL approximation makes a large error in the ionization state of the gas after the overlap, though the effect is still somewhat less than the one found by \citet{newrei:bsv15}. Hence, this test emphasizes the importance of using the correct speed of light in the background equation, and may also serve as an explanation for the difference between CROC and Illustris reionization models.

In conclusion, the RSL approximation, when used correctly, remains a robust and accurate numerical trick to lower the computational expense of an explicit moment-based radiative transfer solver. It does break during the initial stages of a rapidly expanding ionized bubble, as has been shown by \citet{newrei:rba13}, but even that test is artificial - a strong source is not switching on suddenly in a perfectly neutral IGM. 

The process of cosmic reionization is driven by the gradual increase in the production of ionizing photons, and a size of the ionized bubble is determined by the total amount of ionizing photons produced inside - the primary reason why numerous semi-analytical model based on the barrier crossing formalism of \citet{reisam:fzh04} work so well. Thus, the rate of the propagation of ionization fronts in the bulk of the IGM is controlled by the photon production rate in the sources, not by the photon propagation speed.

One application where the RSL approximation may indeed fail is a rapid turn on of a bright quasar. Even for that case the failure is not obvious, as a quasar turns on in a pre-existing ionized bubble, but any RSL-based code used for modeling that process needs to be specifically tested in a manner similar to the test shown in figure \ref{fig:rsl}.

\acknowledgements

I am grateful to Volker Springel for extensive comments and frank discussion. This work was motivated by the discussion at the ``Cosmic Reionization'' program at the Munich Institute for Astro- and Particle Physics (MIAPP) of the DFG cluster of excellence "Origin and Structure of the Universe". Fermilab is operated by Fermi Research Alliance, LLC, under Contract No.~DE-AC02-07CH11359 with the United States Department of Energy. CROC simulations have been performed on the University of Chicago Research Computing Center cluster ``Midway'', on National Energy Research Supercomputing Center (NERSC) supercomputers ``Cori'' and ``Edison'', and on the Argonne Leadership Computing Facility supercomputer ``Mira''. An award of computer time was provided by the Innovative and Novel Computational Impact on Theory and Experiment (INCITE) program. This research used resources of the Argonne Leadership Computing Facility, which is a DOE Office of Science User Facility supported under Contract DE-AC02-06CH11357.

\bibliographystyle{apj}
\bibliography{ng-bibs/self,ng-bibs/rei,ng-bibs/igm,ng-bibs/gals,ng-bibs/misc,ng-bibs/sims,ng-bibs/reisam,ng-bibs/newrei}

\begin{thebibliography}{9}
\expandafter\ifx\csname natexlab\endcsname\relax\def\natexlab#1{#1}\fi

\bibitem[{{Bauer} {et~al.}(2015){Bauer}, {Springel}, {Vogelsberger}, {Genel},
  {Torrey}, {Sijacki}, {Nelson}, \& {Hernquist}}]{newrei:bsv15}
{Bauer}, A., {Springel}, V., {Vogelsberger}, M., {Genel}, S., {Torrey}, P.,
  {Sijacki}, D., {Nelson}, D., \& {Hernquist}, L. 2015, ArXiv e-prints

\bibitem[{{Bouwens} {et~al.}(2015){Bouwens}, {Illingworth}, {Oesch}, {Trenti},
  {Labb{\'e}}, {Bradley}, {Carollo}, {van Dokkum}, {Gonzalez}, {Holwerda},
  {Franx}, {Spitler}, {Smit}, \& {Magee}}]{rei:biot15}
{Bouwens}, R.~J., {Illingworth}, G.~D., {Oesch}, P.~A., {Trenti}, M.,
  {Labb{\'e}}, I., {Bradley}, L., {Carollo}, M., {van Dokkum}, P.~G.,
  {Gonzalez}, V., {Holwerda}, B., {Franx}, M., {Spitler}, L., {Smit}, R., \&
  {Magee}, D. 2015, \apj, 803, 34

\bibitem[{{Furlanetto} {et~al.}(2004){Furlanetto}, {Zaldarriaga}, \&
  {Hernquist}}]{reisam:fzh04}
{Furlanetto}, S.~R., {Zaldarriaga}, M., \& {Hernquist}, L. 2004, \apj, 613, 1

\bibitem[{{Gnedin}(2014)}]{ng:g14}
{Gnedin}, N.~Y. 2014, \apj, 793, 29

\bibitem[{{Gnedin} \& {Abel}(2001)}]{ng:ga01}
{Gnedin}, N.~Y. \& {Abel}, T. 2001, New Astronomy, 6, 437

\bibitem[{{Kravtsov}(1999)}]{misc:k99}
{Kravtsov}, A.~V. 1999, PhD thesis, New Mexico State University

\bibitem[{{Kravtsov} {et~al.}(2002){Kravtsov}, {Klypin}, \&
  {Hoffman}}]{misc:kkh02}
{Kravtsov}, A.~V., {Klypin}, A., \& {Hoffman}, Y. 2002, \apj, 571, 563

\bibitem[{{Rosdahl} {et~al.}(2013){Rosdahl}, {Blaizot}, {Aubert}, {Stranex}, \&
  {Teyssier}}]{newrei:rba13}
{Rosdahl}, J., {Blaizot}, J., {Aubert}, D., {Stranex}, T., \& {Teyssier}, R.
  2013, \mnras, 436, 2188

\bibitem[{{Rudd} {et~al.}(2008){Rudd}, {Zentner}, \& {Kravtsov}}]{sims:rzk08}
{Rudd}, D.~H., {Zentner}, A.~R., \& {Kravtsov}, A.~V. 2008, \apj, 672, 19

\end{thebibliography}

\end{document}